\documentclass[twocolumn,showpacs,aps,prl,superscriptaddress]{revtex4}

\usepackage{graphicx}
\usepackage{amsmath}

\begin{document}

\title{Aharonov-Bohm interference in topological insulator nanoribbons}

\author{Hailin Peng}
\affiliation{Department of Materials Science and Engineering,
Stanford University, Stanford, CA 94305, USA}
\author{Keji Lai}
\affiliation{Department of Applied Physics and Geballe Laboratory
for Advanced Materials, Stanford University, Stanford, CA 94305}
\author{Desheng Kong}
\affiliation{Department of Materials Science and Engineering,
Stanford University, Stanford, CA 94305, USA}
\author{Stefan Meister}
\affiliation{Department of Materials Science and Engineering,
Stanford University, Stanford, CA 94305, USA}
\author{Yulin Chen}
\affiliation{Department of Applied Physics and Geballe Laboratory
for Advanced Materials, Stanford University, Stanford, CA 94305}
\affiliation{Department of Physics, Stanford University, Stanford,
CA 94305, USA}
\affiliation{Stanford Institute for Materials and
Energy Sciences, SLAC National Accelerator Laboratory, 2575 Sand
Hill Road, Menlo Park, CA 94025, USA}
\author{Xiao-Liang Qi}
\affiliation{Department of Physics, Stanford University, Stanford,
CA 94305, USA}
\affiliation{Stanford Institute for Materials and
Energy Sciences, SLAC National Accelerator Laboratory, 2575 Sand
Hill Road, Menlo Park, CA 94025, USA}
\author{Shou-Cheng Zhang}
\affiliation{Department of Physics, Stanford University, Stanford,
CA 94305, USA}
\affiliation{Stanford Institute for Materials and
Energy Sciences, SLAC National Accelerator Laboratory, 2575 Sand
Hill Road, Menlo Park, CA 94025, USA}
\author{Zhi-Xun Shen}
\affiliation{Department of Applied Physics and Geballe Laboratory
for Advanced Materials, Stanford University, Stanford, CA 94305}
\affiliation{Department of Physics, Stanford University, Stanford,
CA 94305, USA}
\affiliation{Stanford Institute for Materials and
Energy Sciences, SLAC National Accelerator Laboratory, 2575 Sand
Hill Road, Menlo Park, CA 94025, USA}
\author{Yi Cui}
\affiliation{Department of Materials Science and Engineering,
Stanford University, Stanford, CA 94305, USA}

\date{\today}

\begin{abstract}

Topological insulators represent novel phases of quantum matter with
an insulating bulk gap and gapless edges or surface states. The
two-dimensional topological insulator phase was predicted in HgTe
quantum wells and confirmed by transport measurements. Recently,
Bi$_2$Se$_3$ and related materials have been proposed as
three-dimensional topological insulators with a single Dirac cone on
the surface and verified by angle-resolved photoemission
spectroscopy experiments. Here, we show unambiguous transport
evidence of topological surface states through periodic quantum
interference effects in layered single-crystalline Bi$_2$Se$_3$
nanoribbons. Pronounced Aharonov-Bohm oscillations in the
magnetoresistance clearly demonstrate the coverage of
two-dimensional electrons on the entire surface, as expected from
the topological nature of the surface states. The dominance of the
primary $h/e$ oscillation and its temperature dependence demonstrate
the robustness of these electronic states. Our results suggest that
topological insulator nanoribbons afford novel promising materials
for future spintronic devices at room temperature.

\end{abstract}
\maketitle

Electronic properties at the surface of a solid can be very
different from those in the bulk. Dangling bonds or reconstruction
of the atoms due to the inevitable loss of periodic lattice
structure could result in surface states absent in the bulk energy
spectrum. In most materials, details of the surface geometry and
chemistry can easily alter these fragile states. Recent theoretical
work, however, has predicted a new class of quantum matter with an
insulating bulk gap and gapless edge or surface states: the
topological insulators in two \cite{1Bernevig, 2Bernevig, 3Kane} and
three dimensions \cite{4Fu, 5Moore, 6Qi}, respectively. These robust
low-dimensional conducting states are topologically protected
against all time-reversal-invariant perturbations, such as
scattering by non-magnetic impurities, crystalline defects, and
distortion of the surface itself. Experimentally, the
two-dimensional (2D) topological insulator phase has been predicted
and realized in HgTe quantum wells \cite{1Bernevig, 7Konig}. The
three-dimensional (3D) topological insulator phase was reported in
Bi$_{1-x}$Sb$_x$ alloy \cite{8Hsieh, 9Hsieh} with complicated
topological surface states. 3D topological insulators with the
simplest possible surface states consisting of a single Dirac cone,
on the other hand, have been proposed theoretically \cite{10Zhang}
in stoichiometric compounds Bi$_2$Se$_3$, Bi$_2$Te$_3$ and
Sb$_2$Te$_3$, and topological surface states have been observed
independently by angle-resolved photoemission spectroscopy (ARPES)
experiments in Bi$_2$Se$_3$ \cite{11Xia}. Recently, the bulk
insulating phase has been achieved in Bi$_2$Te$_3$ \cite{12Chen} as
observed by ARPES. Among these materials, Bi$_2$Se$_3$, a narrow gap
semiconductor for IR detectors and thermoelectric applications
\cite{13Mishra}, has a simple band structure with a single Dirac
cone on the surface and a large non-trivial bulk gap of 0.3eV
\cite{10Zhang, 11Xia}. These properties make Bi$_2$Se$_3$ ideal for
the realization of interesting topological phenomena, such as the
image monopole effect \cite{6Qi, 14Qi} and Majorana fermions
\cite{15Fu}, as well as future room-temperature spintronic
applications.

However, the surface states in topological insulators have been
mainly investigated by ARPES \cite{8Hsieh, 9Hsieh, 11Xia, 12Chen}.
Transport measurements, on the other hand, should be a
straightforward probe to study the properties of such
low-dimensional electronic states \cite{16Fu}. For instance, if
conduction occurs mainly through the 2D channel, the conductance
would scale with geometry of the sample surface rather than that of
the bulk. Under strong magnetic fields, the characteristics of the
2D Fermi surface can be mapped out by the Shubnikov-de Haas (SdH)
oscillations of the magnetoresistance \cite{17Ando}, in which the
magnetoresistance varies periodically with the inverse magnetic
field ($1/B$). Finally, quantum interference effects associated with
the surface states may occur for mesoscopic samples where the low
temperature phase coherence length is comparable to the sample
dimensions \cite{18Aronov}. However, despite extensive transport
experiments on bulk Bi$_2$Se$_3$ since the 1970's \cite{19Kohler},
there has been no report of a conducting surface layer, and the
predicted topological features have not been addressed. Such a
seeming discrepancy is understandable in a macroscopic bulk crystal
since the residual bulk carriers due to crystal defects or thermal
excitations in a small bulk gap semiconductor can easily mask the
transport signatures of the 2D surface electrons.

The contribution of the bulk carriers can be suppressed by reducing
sample size. Quasi-1D nanoribbons, with their large
surface-to-volume ratios, provide excellent geometries for probing
the transport properties of surface states. In this work,
single-crystalline Bi$_2$Se$_3$ nanoribbons are synthesized via a
gold-catalyzed vapor-liquid-solid (VLS) growth \cite{20Morales}. The
layered Bi2Se3 has a rhombohedral phase with the space group
D$^5$$_{3d}$ (R-3m)\cite{21Wyckoff}, and consists of planar,
covalently bonded sheets linked predominantly by van der Waals
interactions (Fig. 1a). Figure 1b shows a typical scanning electron
microscopy (SEM) image of the as-grown Bi$_2$Se$_3$ nanoribbons with
thicknesses of 25-100nm and widths ranging from 50nm to several
micrometers. The nanoribbon has smooth sidewalls and flat surfaces
(Fig. 1c). The presence of a gold nanoparticle at the end of each
nanoribbon suggests the VLS growth mechanism, in which catalyst
particles promote nucleation and unidirectional growth of layered
structures (Figs 1d and 1e). Energy-dispersive X-ray spectroscopy
(EDX) analysis reveals uniform chemical composition along the ribbon
length with a Bi:Se atomic ratio of 2:3, indicating stoichiometric
Bi$_2$Se$_3$ free of detectable impurities. The nanoribbons are
single crystalline rhombohedral phase with atomically smooth edges,
length along the [11-20] direction, and height parallel to the
c-axis, as shown in transmission electron microscopy, the selected
area electron diffraction pattern (Fig. 1f, inset), and the X-ray
diffraction (XRD) pattern (Fig. 1g). By controlling the growth
temperature, using Sn/Au alloy as the catalyst, or post-annealing in
Se vapor, we are able to vary the electron density without changing
the crystal structure. The total electron density measured by the
Hall effect is between $3\times10^{13}$ to $2\times10^{14}$
cm$^{-2}$, with an effective Hall mobility $\sim 2\times10^3$
cm$^2$/Vs extracted from the sheet resistance . Such a high electron
density indicates the existence of bulk carriers in our nanoribbons,
while their contribution is suppressed compared with macroscopic
samples. Results reported below are only observed in samples with
relatively low total density $\sim 5\times10^{13}$ cm$^{-2}$.

\begin{figure}[!t]
\begin{center}
\includegraphics[width=0.9\columnwidth]{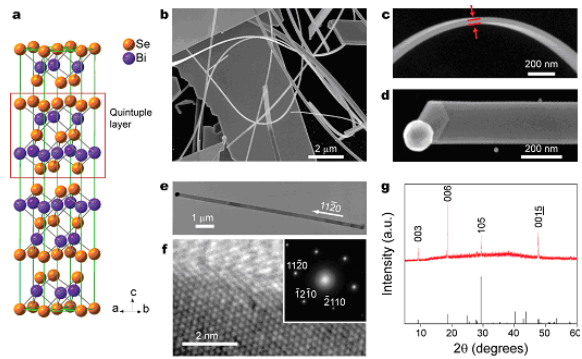}
\end{center}
\caption{\label{1} $\bf a$. Layered crystal structure of
Bi$_2$Se$_3$ with quintuple layers ordered in the Se-Bi-Se-Bi-Se
sequence along the $c$ axis. $\bf b$. SEM image of as-grown
nanoribbons from Bi$_2$Se$_3$ evaporation by VLS growth. $\bf c$.
High-resolution SEM image showing the ribbon shape and flat surfaces
of a 30nm-thick nanoribbon. $\bf d$. Close-up view of a nanoribbon
with a gold nanoparticle catalyst at the end. $\bf e$. TEM image of
a Bi$_2$Se$_3$ nanoribbon with a nanoparticle at the tip. $\bf f$.
High-resolution TEM image of the edge of the Bi$_2$Se$_3$ nanoribbon
showing the smooth surfaces with little ($<$1 nm) amorphous layers.
The selected area electron diffraction pattern (inset) indicates
that the ribbon is single-crystalline all along its length. The
growth direction of nanoribbon is along [11-20]. $\bf g$. X-ray
powder diffraction pattern of Bi$_2$Se$_3$ nanoribbons with a
reference diffractogram. The products were identified as the
single-phase rhombohedral Bi$_2$Se$_3$ (R-3m, $a = b$ = 0.4140 nm,
and $c$ = 2.8636 nm).}
\end{figure}

The quasi-1D narrow nanoribbons in Fig. 1e are model systems for
studying the topological surface states. When an external magnetic
field ($B$) is applied along the length of the nanoribbon, quantum
interference effects will occur if the conduction electrons remain
phase coherent after completing closed trajectories, each encircling
a certain magnetic flux. For the bulk carriers, there exist various
sample-specific, or more precisely, impurity-dependent loops,
resulting in universal conductance fluctuations (UCF) \cite{22Lee}
with aperiodic field dependence, commonly observed in small metallic
and semiconducting structures. For the 2D states covering the entire
surface, however, all phase coherent trajectories participating in
the quantum interference enclose the same area perpendicular to the
$B$-field (Fig. 2a). The low temperature transport therefore
exhibits periodic magnetoresistance oscillations, a hallmark of the
well-known Aharonov-Bohm (AB) effect \cite{23Aharonov}, with a
characteristic period of the external magnetic field $\Delta$$B =
\Phi_0 / S$, where $\Phi_0 = h/e$ is the flux quantum, $S$ the cross
sectional area of the nanoribbon, $h$ the Planck's constant and $e$
the electron charge, respectively.

\begin{figure}[!t]
\begin{center}
\includegraphics[width=0.8\columnwidth]{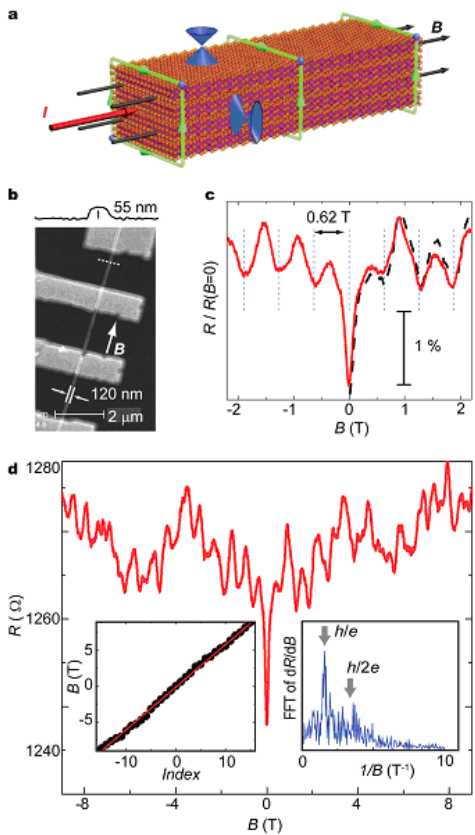}
\end{center}
\caption{\label{2} $\bf a$. Schematic of a 2D topological surface
states of a layered Bi$_2$Se$_3$ nanoribbon under a magnetic field
along the ribbon length. The red and black arrows correspond to the
electric current and magnetic field lines, respectively. The two
cones on the top and side surfaces illustrate the Dirac surface
states propagating on all surfaces with linear dispersion. The green
loops encircling the same magnetic flux stand for phase coherent
paths through which the surface electrons interfere. $\bf b$. SEM
image of a Bi$_2$Se$_3$ nanoribbon, 120nm in width, contacted by
four Ti$/$Au electrodes. The thickness of the nanoribbon is measured
by AFM (a line cut in the inset) to be 55nm. $\bf c$. Normalized
magnetoresistance of the nanoribbon in radial magnetic fields at 2K.
A clear modulation of the resistance with a period of 0.62T is
observed, corresponding to one flux quantum ($h/e$) threaded into
the cross section of the nanoribbon. The solid red trace (up sweep)
was taken with a scan rate of 3mT$/$sec and the dashed black line
(down sweep) at 10mT$/$sec. $\bf d$. Magnetoresistance in the full
field range of $\pm$9T. Inset on the left, magnetic field position
of well developed resistance minima. Inset on the right, Fast
Fourier transform (FFT) of the derivative d$R$/d$B$ in the entire
field range. Locations of $h/e$ and $h/2e$ flux quantization are
labeled.}
\end{figure}

To probe the quantum interference effect in the Bi$_2$Se$_3$
nanoribbons, we fabricated four-point probe devices by depositing
Ti/Au Ohmic contacts. A representative device is shown in Fig. 2b,
where the width $w$ and thickness $t$ of the ribbon are determined
by SEM and AFM, respectively. The four-terminal magnetoresistance is
measured in a 9-Tesla Quantum Design PPMS system with a base
temperature of 2K. At low magnetic fields $|B| < 0.15$T, the weak
anti-localization effect \cite{24Hikami} with a sharp cusp at zero
field is seen in Fig. 2c, consistent with the presence of strong
spin-orbit coupling in Bi2Se3. For fields 0.15T $< |B| <$ 2T,
pronounced and reproducible resistance oscillations with a period of
0.62T are observed. For the $h/e$ AB effect, this period is
associated with an area of $6.7\times10^{-15}$ m$^2$, in excellent
agreement with the measured cross sectional area of the nanoribbon
$S = w\times t = 6.6\times10^{-15}$ m$^2$ (Fig. 2b). At higher
fields (Fig. 2d), the perfect periodicity deteriorates slightly, but
a roughly linear relation of 0.6T per resistance minimum still holds
up to 9T. As shown in the inset of Fig. 2d, the prominent $h/e$
oscillation and a weak second harmonic at $h/2e$ are better seen by
taking the fast Fourier transform (FFT) of the derivative d$R$/d$B$,
a method commonly applied to separate the oscillatory part from
slow-varying background \cite{25Huber}. The noisy background in the
FFT spectrum and the deviation from pure $h/e$ periodicity at high
fields are presumably due to the contribution from the bulk states,
which superimpose aperiodic UCF onto the clean AB oscillations. We
also note that SdH oscillations of the bulk electrons may be
responsible for the slow-varying background seen in Fig. 2d. Since
quantum confinement in the plane perpendicular to the field
direction is not negligible, several electric sub-bands may form in
our nanoribbons, and magnetic depopulation of these sub-bands would
modify the usual $1/B$ relation at high Landau level fillings
\cite{26Berggren}. Nevertheless, neither of these bulk-related
effects could result in B-periodicity over the large field range in
this experiment. Therefore, the AB effect in Bi$_2$Se$_3$
nanoribbons indisputably manifests the existence of the conducting
surface states, which, based on theories and the ARPES data
\cite{10Zhang, 11Xia}, are highly exotic as opposed to the trivial
surface inversion effect in certain semiconducting materials.

The observation of the AB oscillations in Bi$_2$Se$_3$
nanostructures provides important insights into the topological
surface states. First of all, although a significant portion of the
conduction is carried by the bulk states, the interaction between
bulk and surface electrons does not destroy the phase coherence of
the surface states. The oscillation amplitude observed in our
experiment is on the order of the quantized conductance ($e^2/h$),
comparable to other quasi-1D nanostructures \cite{25Huber} and
disordered metal cylinders \cite{18Aronov}. Since full revolution
around the perimeter of the ribbon is required for the interference
effect, our result demonstrates that the surface states not only
exist on the top and bottom (0001) surfaces but also propagate
coherently through the side walls of the nanoribbon. Given that the
side surface is terminated with dangling bonds and adjacent quasi-2D
layers are predominantly linked by van der Waals coupling, the
coherent motion of the electrons through the side surface is highly
nontrivial, providing further evidence for the topological
robustness of the surface states. Finally, the primary oscillation
period corresponds to the Aharonov-Bohm $h/e$ quantization, while
the Altshuler-Aronov-Spivak (AAS) oscillation with $h/2e$ period
\cite{27Altshuler, 28Bachtold} originating from weak (anti)
localization is suppressed. The absence of the AAS effect, at least
in the low magnetic field region, may be related to the distinct
property of the topological surface states, i.e., electrons with
opposite momentum have opposite spin polarization. For these states,
back scattering events are forbidden for $B=0$ but are present for
finite field strengths even if the flux of the field is an integer
multiple of $h/2e$ flux quanta. Consequently, the anti-localization
behavior observed at $B = 0$ is absent for other integer multiples
of $h/2e$ flux, leading to the suppression of the AAS effect.

\begin{figure}[!t]
\begin{center}
\includegraphics[width=0.8\columnwidth]{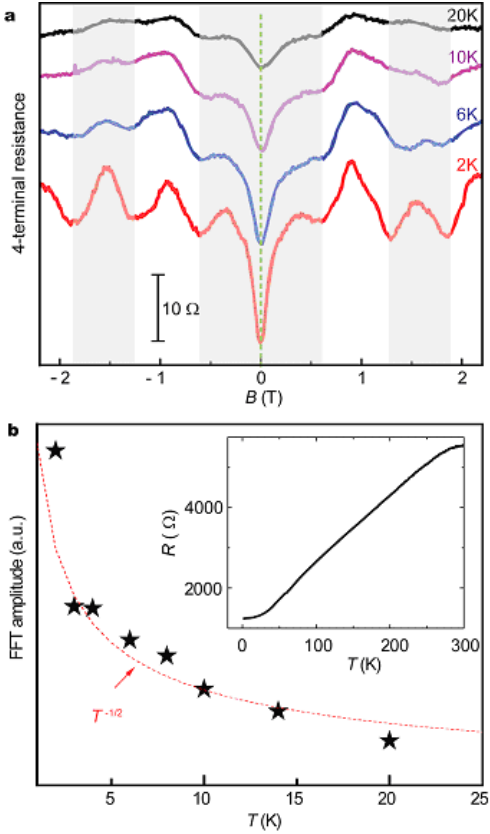}
\end{center}
\caption{\label{3} $\bf a$. Magnetoresistance vs magnetic field at
four different temperatures. The curves are vertically displaced for
clarity. $\bf b$. T-dependence of the FFT amplitude of the primary
$h/e$ oscillations. The data roughly follow the expected $T^{-1/2}$
relation (dashed line) between 2K and 20K. Inset. Zero field cooling
curve from 300K to 2K.}
\end{figure}

Further information about the AB effect can be obtained by the
temperature evolution of the magnetoresistance oscillations. As
shown in Fig. 3a, most oscillatory features persist up to 20K,
beyond which the zero field resistance increases considerably (inset
of Fig. 3b), presumably due to phonon scattering. Again, we take the
FFT of dR/dB to analyze the T-dependence data. In Fig. 3b, the $h/e$
oscillation amplitude roughly scales with $T^{-1/2}$ between 2K and
20K, where the inelastic phonon scattering is negligible. The same
T-dependence was observed in mesoscopic metal rings and explained by
the averaging of conduction channels in the absence of inelastic
scattering \cite{29Washburn}. At high densities with weak
electron-electron interaction, the phase-breaking time $\tau_\phi$
can be estimated from the thermal broadening of the Fermi-Dirac
distribution, or $\tau_\phi \sim \hbar / k_BT$. As a result, the
oscillation amplitude $\Delta R$ scales with the phase coherent
diffusion length $L_\phi = (D\tau_\phi)^{1/2} \sim T^{-1/2}$, where
$D$ is the diffusion constant, in accordance with the experimental
data.

\begin{figure}[!t]
\begin{center}
\includegraphics[width=0.8\columnwidth]{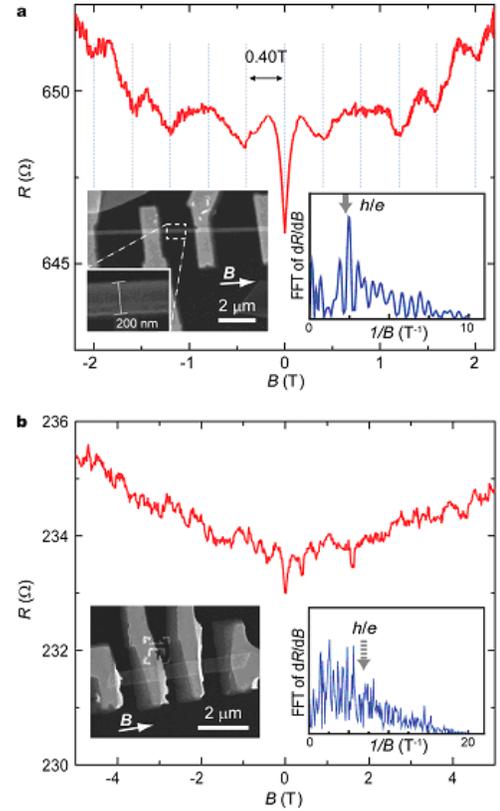}
\end{center}
\caption{\label{4} $\bf a$. AB oscillations of the nanoribbon shown
in the inset. The label of $h/e$ marks the calculated (from the
cross section 200nm $\times$ 50nm) $1/B$ frequency in the FFT plot.
$\bf b$. Aperiodic magnetoresistance of a wide ribbon (570nm
$\times$ 50nm) and the SEM in the inset. The FFT spectrum shows no
apparent peaks. The location of $h/e$ is again indicated for
comparison.}
\end{figure}

In the end, we briefly discuss the quantum interference effect in
two more Bi$_2$Se$_3$ nanoribbon devices at 2K. For the relatively
narrow ribbon in Fig. 4a, well defined resistance minima are
observed at multiples of 0.4T except for weak shoulders at $\sim$
0.8T. This period is better seen in the FFT plot. The inferred area
from the $h/e$ flux quantization again matches the cross sectional
area (200nm $\times$ 50nm) well within the measurement error bars.
On the other hand, the magnetoresistance of the wide ribbon (570nm
$\times$ 50nm) in Fig. 4b does not display obvious periodicity, and
the characteristics of its Fourier spectrum are reminiscent of the
UCF effect. In other words, neither the bulk nor the surface
electrons can produce a coherent path that predominantly encloses a
fixed amount of flux. Thus the phase coherent length $L_\phi$ in our
nanoribbons is likely to be around 0.5$\mu$m at 2K, consistent with
an order-of-magnitude estimate from SdH measurements. Further
experiments down to lower temperatures are expected to enhance
$L_\phi$ for the observation of AB effects in wider ribbons and will
be investigated in the future.

In conclusion, we have fabricated topological insulator materials in
nanoribbon form, with large surface-to-volume ratios. We report the
first transport measurement in this class of materials, based on the
AB oscillations of the conductance associated with the topological
surface states. The distribution of the AB periods provides the
clear evidence to distinguish the surface conductance from the bulk,
paving the way to eventually eliminate the bulk conduction.

\end{document}